\documentclass[conference]{IEEEtran}
\makeatletter
\def\ps@headings{%
\def\@oddhead{\mbox{}\scriptsize\rightmark \hfil \thepage}%
\def\@evenhead{\scriptsize\thepage \hfil\leftmark\mbox{}}%
\def\@oddfoot{}%
\def\@evenfoot{}}
\makeatother
\usepackage{verbatim}  
\usepackage{graphicx}
\usepackage[tight]{subfigure}
\usepackage{amsmath, amsthm, amssymb}

\usepackage{algorithm}
\usepackage[noend]{algorithmic}
\usepackage{tweaklist}   

\IfFileExists{url.sty}{\usepackage{url}}
                      {\newcommand{\url}{\texttt}}
\newcommand{\urlsamefont}[1]
{
\urlstyle{same}\url{#1}
}

\newcommand{\FR}{free-rider}
\newcommand{\BT}{BitTorrent}

\usepackage{cite}

\begin{document}
\title{Reinforcement Learning in BitTorrent Systems}

\author{
Rafit Izhak-Ratzin *&& Hyunggon Park *&& Mihaela van der Schaar\\
PaloAlto && Electrical Engineering Institute && Electrical Engineering Department\\ 
Networks && Ewha Womans University && University of California\\
Sunnyvale, CA && Seoul, Korea && Los Angeles, CA
}
\maketitle
\begin{abstract}
Recent research efforts have shown that the popular BitTorrent protocol does 
not provide fair resource reciprocation and may allow free-riding. 
In this paper, we propose a BitTorrent-like protocol that replaces the peer
selection mechanisms in the regular BitTorrent protocol with a novel 
reinforcement learning based mechanism.

Due to the inherent opration of P2P systems, which involves repeated 
interactions among peers over a long period of time, the peers can efficiently 
identify free-riders as well as desirable collaborators by learning the 
behavior of their associated peers.
Thus, it can help peers improve their download rates and discourage 
free-riding, while improving fairness in the system. 

We model the peers' interactions in the BitTorrent-like network as a repeated 
interaction game, where we explicitly consider the 
strategic behavior of the peers. 
A peer, which applies the reinforcement learning based mechanism, uses a 
partial history of the observations on associated peers' statistical reciprocal 
behaviors to determine its best responses and estimate the corresponding 
impact on its expected utility.  
The policy determines the peer's resource reciprocations with other 
peers, which would maximize the peer's long-term performance, thereby making 
foresighted decisions.

We have implemented the proposed reinforcement-learning based mechanism and 
incorporated it into an existing BitTorrent client. 
We have performed extensive experiments on a controlled Planetlab test bed. 
Our results confirm that our proposed protocol 
(1) promotes fairness in terms of incentives to each peer's contribution e.g. 
high capacity peers improve their download completion time by up to 33\%, 
(2) improves the system stability and robustness e.g. reducing the peer 
selection fluctuations by 57\%, 
and (3) discourages free-riding e.g. peers reduce by 64\% their upload to \FR, 
in comparison to the regular \BT~protocol.

\end{abstract}


\footnote{* This work was done while Dr. Izhak-Ratzin and Dr.Park were at UCLA}

\section{Introduction}
\label{sec:introduction}

Peer-to-peer (P2P) content sharing protocols dominate the traffic on the 
Internet, and thus, have become an important part in building scalable 
Internet applications~\cite{ipoque}. The P2P protocols are used by a variety 
of Internet applications such as content distribution~\cite{BitTorrent}, voice 
over IP~\cite{Skype}, and streaming multimedia P2P 
applications~\cite{PPLive, UUSEE}. 
%

In P2P content distribution systems, fairness among peers is an important 
factor, as it encourages peers to actively collaborate in disseminating 
content, which can lead to an improved system performance. 
However, even BitTorrent~\cite{cohen03}, one of the most popular protocols used 
in P2P content distribution, does not provide fair resource reciprocation,
particularly for node populations having heterogeneous upload 
bandwidths~\cite{piatek07, bharambe06, guo05,legout07}. 
This is because the tit-for-tat strategy implemented in BitTorrent only 
exploits a short-term history for making upload decisions. More specifically, 
upload decisions are made based on the most recent observations of the resource 
reciprocation. This also implies that the upload decisions are short 
backward-looking and not forward-looking, i.e., the decisions are not 
foresighted.  
Thus, a peer can keep following the tit-for-tat policy only if it continuously 
uploads pieces of a particular file to its associated peers and as long as it 
receives pieces of interest in return.
However, this is not always feasible as irrespective of peers' willingness to 
cooperate, they may not always have pieces in which the other peers are 
interested in~\cite{piatek08}. However, such behavior is still perceived as a 
lack of cooperation for interacting peers. In addition, it has been shown that 
BitTorrent systems do not effectively cope with selfish peers' behaviors such 
as free-riding~\cite{liogkas06, locher06, sirivianos07}, because of their 
built-in optimistic unchoke mechanism. 
While the optimistic unchoke mechanism enables peers to continuously discover 
better peers (or leechers) to reciprocate resources, it can provide a major 
opportunity for selfish peers to obtain data without uploading in return.
This mechanism may also lead to unfairness in the system, as it forces 
high-capacity peers to interact with low-capacity peers.

Unlike the approaches that are using short-term observation history, 
reputation-based schemes have been proposed to overcome the limitations of 
tit-for-tat and optimistic unchoke mechanisms by exploiting global histories 
(e.g.,~\cite{buchegger04, xiong04, yang05}). 
However, in order to maintain such a global history across peers, 
these approaches require significant communication overhead.   
Moreover, the reliability of global history can be unclear as peers may exhibit 
different reciprocation behaviors with different peers.
%
Alternatively, the long-term local (or private) history of upload behaviors 
with associated peers' is used in several other reputation-based approaches 
such as~\cite{eDonkey, piatek08, Buddies, Teams}. 
While these approaches can reduce the communication overheads, the focus of 
these systems is still on maximizing the \emph{immediate} utility, which may be 
less desirable than maximizing the \emph{long-term} utility, as peers can 
repeatedly interact with each other over a long period of time.

In this paper, we model the peer interactions in the BitTorrent-like network 
as a repeated interaction game -- repeated interactions (i.e., reciprocating 
resources) among several participants (i.e., peers) in which a participant 
takes actions (i.e. unchoke peers) so as to maximize long term reward (i.e., 
cumulative download rates). 
The underlying state of the environment changes stochastically, and is 
contingent upon the decisions of the participants. 
In our model, peers can apply reinforcement-learning (RL) to make upload 
decisions. We explicitly consider the strategic behaviors of peers, where the 
peers can observe partial historical information about the reciprocation 
behaviors of their associated peers. 
Based on this information, the peers that apply the RL-based strategy can 
estimate their future expected rewards, and then, can determine accordingly 
their best responses. The future expected rewards can be determined using  
various types of interactive learning techniques. 
We use reinforcement learning, since it enables the peers to improve their peer 
selection strategies based solely on the knowledge of their past interactions, 
but not on the knowledge of the complete reciprocation behaviors of the peers 
in the entire network. 
The reinforcement learning enables each peer to forecast the impact of the 
current peer selection on the future expected utility and to maximize 
it.
Therefore, 
the RL-based peer selection mechanism replaces both the tit-for-tat and the 
optimistic unchoke mechanisms in the regular BitTorrent protocol.

Note that our protocol supports a non real-time media transmission scenario, 
which has received less attention in the multimedia research community compared 
to the on-demand media streaming scenario. In this type of protocols, the 
requested content needs to be completely downloaded before it is displayed. 
Thus, the ordering of which pieces are downloaded first is not important, but 
the overall time required for completely downloading the content is important.
Note that, however, the proposed protocol can be easily adapted to on-demand media streaming 
applications using existing techniques such as~\cite{BASS05, BiToS06, choe07}.

The proposed protocol consists of three main processes:
\begin{itemize}
\item \emph{Learning Process}, which provides updated information about 
statistical behaviors of the associated peers' resource reciprocation, 
\item \emph{Policy Finding Process}, which computes the peer selection policy 
based on the reinforcement learning, and
\item \emph{Decision Process}, which determines the associated peers that will 
be unchoked and choked during every rechoke period based on the 
peer selection policy. 
\end{itemize}
%
We implemented our proposed protocol on top of an actual BitTorrent client, and 
performed extensive experiments in a controlled Planetlab test bed.
The new proposed algorithm is executed simply through policy modifications to 
existing clients with no changes to the BitTorrent protocol.
Our protocol does not demand full adoption or sparse adoption of the RL-based 
peer selection mechanism (as in~\cite{piatek07}) and can be run by any number 
of peers in a BitTorrent-like network. 
We evaluated and quantified the performance of the proposed protocol, and 
compared its performance with the regular BitTorrent protocol.
Based on the experimental results, the proposed protocol provides the following
advantages against the regular BitTorrent protocol: 
\begin{enumerate}
\item It discourages free-riding by limiting the upload to non-cooperative 
peers.
\item It promotes cooperation among high-capacity peers. 
\item It improves fairness; the peers that contribute more resources 
      (i.e., higher upload capacities) can achieve higher download rates. 
      While, the peers that contribute fewer resources may achieve lower 
      download rates. 
\item It improves the system robustness by minimizing the impact of free-riding 
      on the contributing peers' performance.
\item It improves the stability of the peer selection mechanism, which affects 
      directly the performance of the system.
\end{enumerate}

The rest of the paper is organized as follows. In Section~\ref{sec:BToverview}, 
we briefly describe the BitTorrent systems. In Section~\ref{sec:MDP}, we 
briefly define the game and the adopted reinforcement learning solution and 
describe the RL-based 
peer selection mechanism. Section \ref{sec:design} presents the design of the 
proposed protocol. Details of our protocol implementation are discussed in 
Section~\ref{sec:implementation}. The experimental results are presented in 
Section~\ref{sec:evaluation}. Finally, we discuss related work in 
Section~\ref{sec:related}, and the conclusions are drawn in 
Section~\ref{sec:conclusion}.

\section{BitTorrent Overview}
\label{sec:BToverview}
In this section we briefly overview the BitTorrent protocol~\cite{cohen03}.
The BitTorrent protocol is often adopted for P2P content distribution, because 
it can efficiently scale with a large number of participating clients. 

Before the content distribution process begins, the content provider divides 
the possessed data content into multiple \textit{pieces}, or \emph{chunks}. 
Then, the provider creates a \textit{metainfo file}, which contains information 
necessary to initiate the content downloading process. The metainfo file 
includes the address of the \textit{tracker}, which plays the role of 
coordinator that facilitates peer discovery.
A client downloads the metainfo file before joining a \textit{torrent} 
(or \emph{swarm}) -- a group of peers interested in a particular content. 
Then, it connects the tracker to receive a \textit{peer set}, which consists of 
randomly selected peers currently exchanging the same content. 
The peer set may include both \textit{leechers}, peers that are still 
downloading content pieces, and \textit{seeds}, peers that have the entire 
content and upload it to other peers. 
The client can then connect and exchange (or, \emph{reciprocate}) its content 
pieces with its \textit{associated peers} -- the peers in its peer set.

While reciprocating content pieces, each leecher determines a set of peers 
among its peer set from where it can download its content pieces.
The peer selection is determined by \emph{choking mechanisms} which determined 
the \textit{choking decisions}.
BitTorrent leechers adopt two choking mechanisms: the \textit{tit-for-tat} 
resource reciprocation mechanism and the \textit{optimistic unchoke} mechanism.  
The tit-for-tat mechanism prefers the peers that upload their pieces at the 
highest rate among the associated peers.   
Specifically, every 10 seconds (or \textit{rechoke period}), a leecher 
checks the current download rates from its associated peers and selects the 
peers that are uploading their pieces at the highest rates. Then, the leecher 
uploads only to the selected associated peers, while choking (i.e., blocking 
download) the rest of them during the rechoke period. 

The available upload bandwidth is equally divided into the unchoked peers.
The optimistic unchoke mechanism reserves a portion of the available upload 
bandwidth to provide pieces to peers that are randomly selected.
The purpose of this mechanism is to enable the leechers to continuously 
discover better peers to associate itself with, and bootstrap newly joining 
leechers into the tit-for-tat mechanism.
Optimistic unchokes are randomly rotated among the associated peers, typically 
once every three rechoke periods, allowing enough time for leechers to 
demonstrate their cooperative behaviors.

The number of unchoked peers (slots) may vary depending on specific 
implementation, and it can be fixed or dynamically changed as a function of 
the available upload bandwidth.

Seeds deploy different choking mechanism as they already completed to download 
content. The most common implementation is based on round-robin schedule, 
aiming to distribute data uniformly. This implementation is also deployed in 
our implementation.

\section{Reinforcement Learning for Resource Reciprocation in P2P Networks}
\label{sec:MDP}

Peers in BitTorrent-like systems often make repeated decisions to select 
unchoked peers given their dynamically changing environment. 
The evolution of the peers' interactions across the various rechoke periods is  
modeled as a repeated interaction game. We assume that this stochastic game is 
played over a long period of time, in order to support several popular 
applications such as video streaming or large-size file delivery.  

In each time slot (i.e., rechoke period), every peer is in a state and needs to 
select its optimal action. The peers choose their own actions 
independently and simultaneously in each rechoke period. After that, the peers 
are rewarded for taking their actions and transit into the next states. 
The reward (received by each peer) and the state transition are contingent upon 
other peers' states and actions. 

During the repeated multiple peers' interactions, the peers can only observe a 
partial history of their associated peers' reciprocation behaviors. Based on 
these observations, the peers that adopt the RL-based peer selection policy can 
estimate their future expected rewards and can identify their best responses. 
The estimation of the future expected reward can be computed using different 
types of learning schemes. 
In this paper, we use reinforcement learning~\cite{Hu}, as it allows the 
peers to improve their peer selection strategy using only knowledge of their 
own past reciprocation, without knowing the complete knowledge of reciprocation 
behavior of the associated peers in the network. 

Formally, a reinforcement learning environment can be represented by a tuple, 
$\left\langle \mathbf{I}, {\mathcal{S}},{\mathcal{A}}, P, R   \right\rangle$.
$\mathbf{I}$ is a set of peers in the game. If there are $M$ peers in the game, 
$\mathbf{I}$ can be denoted by $\mathbf{I} = \left\{1, \ldots, M\right\}$. 
${\mathcal{S}}$ is the set of state profiles of all peers in the game, i.e., 
${\mathcal{S}} = \mathbf{S}_1\times \cdots \times \mathbf{S}_M$, where 
$\mathbf{S}_j$ is the state space of peer $j$. ${\mathcal{A}}$ is the 
joint action space 
${\mathcal{A}} = \mathbf{A}_1\times \cdots \times \mathbf{A}_M$, where 
$\mathbf{A}_j$ is the action (peer selection) space for peer $j$.
$P: {\mathcal{S}} \times {\mathcal{A}} \times {\mathcal{S}}  \to [0,1]$ is a 
state transition probability function that maps from state profile 
$\mathcal{S}(t) \in {\mathcal{S}} $ at time $t$ into the next state profile 
$\mathcal{S}(t + 1) \in {\mathcal{S}}$ at time $t+1$ given corresponding joint 
actions $\mathcal{A}(t) \in {\mathcal{A}}$. Note that $t$ here is discrete and 
measured in time slots. 
Finally, $R :\mathcal{S} \times {\mathcal{A}} \rightarrow \mathbb{R}_{+}^M$ is 
a reward vector function defined as a mapping from the state profile 
$\mathcal{S}(t) \in {\mathcal{S}} $ at time $t$, and corresponding joint 
actions $\mathcal{A}(t) \in {\mathcal{A}}$, to a  
vector with each element being the reward to a particular peer.

To find the optimal policy in the game (e.g., a stochastic game 
model~\cite{fudenberg99, shaply53}), peers may require the entire history of 
the interactions among peers in the networks. However, this may be infeasible 
for real P2P networks. Unlike such games, finding a RL-based policy 
only requires the peers' own histories of observations through their 
experiences (or interactions). Therefore we expect the RL-based peer selection 
policy to be suboptimal.

The history of observations in the network up to time $t-1$ is defined as \\
${H}(t) = $
\begin{equation}
\{\mathcal{S}(0), \mathcal{A}(0), \mathcal{R}(0), \ldots,
\mathcal{S}({t-1}), \mathcal{A}({t-1}), \mathcal{R}({t-1}) \} \in \mathcal{H}(t)
\end{equation}
Which summarizes all previous states, actions and rewards of the peers in the 
network up to time $t-1$, where $\mathcal{H}(t)$ is the set of all possible 
histories up to time $t-1$. 
Since a peer $j$ cannot access the entire history of observations, i.e.,
$\mathcal{H}(t)$, but rather a portion of $\mathcal{H}(t)$, a set of 
observations that peer $j$ can access is expressed as 
$\mathbf{O}_j(t) \in \mathcal{O}_j$ and 
$\mathbf{O}_j(t) \subseteq \mathbf{H}(t)$. 
Note that the current state $\mathbf{S}_j(t)$ is always observable, i.e., 
$\mathbf{S}_j(t) \in \mathbf{O}_j(t)$. The state transition probability is 
calculated from $\mathbf{O}_j(t)$. 

\subsubsection{State Space of Peer $j$ -- $\mathbf{S}_j$}
The state of peer $j$ represents the set of resources received from the peers 
in $C_j$, where $C_j$ denotes the set of peers associating with peer $j$.
Thus, it may represent the uploading behavior of its associated peers, or 
equivalently, it can capture peer $j$'s download rates from its associated 
peers. The upload rates from peer $i \in C_j$ to peer $j$ at time $t$ are 
denoted by $L_{ij} (t)$. In our proposed protocol, an uploading behavior of 
peer $i$ observed by peer $j$ is denoted by $s_{ij}$, and defined as
\begin{align}
  s_{ij} = \left\{
\begin{array}{ll}
		1, &\textrm{if $L_{ij}  > \theta_j$ }\\
        0, &\textrm{otherwise}\\
\end{array} \right.
\end{align}
where $\theta_j$ is a pre-determined threshold of peer $j$.\footnote{In order 
to minimize the computational complexity, we consider $s_{ij} \in \{0, 1\}$ in 
this paper. However, the granularity of state can be easily extended.}
Thus, $s_{ij}$ can be expressed with one bit and the state space of peer $j$ 
can be expressed as
\begin{equation}
\mathbf{S}_j  = \left\{ {\left. { (s_{1j}, \ldots, s_{Nj} } )
\right| s_{ij}  \in \{ 0, 1\} \,\, \textrm{for all} \,\, i \in C_j } \right\}
\end{equation}
where $N$ denotes the number of peer $j$'s associated peers in $C_j$, i.e., $|C_j| =
N$. 
Therefore, a state $\mathbf{S}_j(t) \in \mathbf{S}_j$ can capture the uploading 
behavior of the associated peers at time $t$. 

\subsubsection{Action Space of Peer $j$ -- $\mathbf{A}_j$}
The action of peer $j$ represents the set of its peer selection decisions. 
The peer selection decision of peer $j$ to peer $i$ at time $t$ is denoted
by $a_{ji}$, and is defined as 
\begin{align}
		a_{ji}(t) = \left\{
\begin{array}{ll}
		0, &\textrm{if peer $j$ chokes peer $i$ }\\
        1, &\textrm{otherwise}.\\
\end{array} \right.
\end{align}
Thus, $a_{ji}$ can also be expressed with one bit.
The action space of peer $j$ can be expressed as
\begin{equation}
\mathbf{A}_j  = \left\{ {\left. { (a_{j1} , \ldots ,a_{jN} } )
\right| a_{ji}  \in \{ 0, 1\} \,\, \textrm{for all} \,\, i \in C_j } \right\},
\end{equation}
Hence, an action $\mathbf{A}_j(t) \in \mathbf{A}_j$ is of vector that consists 
of peer $j$'s peer selection decisions to its associated peers at time $t$. 
In the proposed protocol, 
we assume that peer $j$ is able to unchoke 
$N_u ( \le N)$ peers.
Note that peer $j$ allocates the same amount of upload bandwidths to all 
unchoked peers, the variable case can be future explored.
Thus, the bandwidth allocated to an unchoked peer $i$ by peer $j$ at time 
$t$ is determined by $L_{ji} (t) = B_j /N_u$, where $B_j$ is the maximum 
upload bandwidth available to peer $j$.   

\subsubsection{State Transition Probability of Peer $j$} 
A state transition probability represents the probability that an action 
$\mathbf{A}_j(t) \in \mathbf{A}_j$ of peer $j$ in state $\mathbf{S}_j(t) \in \mathbf{S}_j$ at time $t$
will lead to another state $\mathbf{S}_j(t+1) \in {\mathbf{S}}_j $
at time $t+1$. This can be expressed as 
\begin{equation}
		P_{\mathbf{A}_j(t)} (\mathbf{S}_j(t), \mathbf{S}_j(t+1) ) = \Pr
		(\mathbf{S}_j(t+1) |\mathbf{S}_j(t), \mathbf{A}_j(t)
	).
\end{equation}
A peer $j$ can estimate the state transition probability functions 
based on its history interactions of $\mathbf{S}_j(t')$, $\mathbf{A}_j(t')$ and
$\mathbf{S}_j(t'+1)$ for $t' < t$, which may 
be stored in a transition table. 
While we deploy an empirical frequency based algorithm to estimate the state 
transition probability function in this paper, which is presented in 
Section~\ref{subsec:TheLearningProcess}, other algorithms 
(e.g.,~\cite{Park2009TMM}) can also be used. 

\subsubsection{The Reward of Peer $j$ -- $R_j$}
The reward of a peer in a state is its total estimated download rate in that 
state. Thus, a reward of a peer in a state is the sum of the estimated download 
rates from all of its associated peers. More specifically, a reward of peer $j$ 
from state $\mathbf{S}_j (t) \in \mathbf{S}_j$ 
can be expressed as 
\begin{equation}
	R_j(\mathbf{S}_j (t)) = \left\langle \mathbf{S}_j, [L_{ij}]_{i \in
	C_j}\right\rangle = \sum_{i \in C_j} L_{ij}
\end{equation}
where $\left\langle \mathbf{X}, \mathbf{Y} \right\rangle$ denotes the 
inner-product between two vectors of $\mathbf{X}$ and $\mathbf{Y}$.
A set of rewards for
all peers in the system is denoted by $\mathcal{R} = \{ R_1, \ldots, R_N \}$. 

\subsubsection{RL-based Policy $\pi_j$}
The policy $\pi_j$, which can be obtained from the reinforcement
learning, can provide a specific action $\mathbf{A}_j(t)$ 
for peer $j$ in state $\mathbf{S}_j(t)$ at time $t$, i.e., 
$\pi_j: \mathcal{S}_j \rightarrow \mathcal{A}_j$. Thus,
$\mathbf{A}_j(t) = \pi_j \left(\mathbf{S}_j(t) \right)$. 

 
The actions that the policy provides to peer $j$ are determined such that they  
can maximize the cumulative discounted expected reward, which is 
defined for a peer $j$ in state $\mathbf{S}_j(t)$ at time $t=t_c$ 
given a discount factor $\gamma_j$ as
\begin{equation}
  R_j^{f}(\mathbf{S}_j(t_c)) \triangleq \sum\limits_{t = t_c  + 1}^\infty
  {\gamma_j ^{(t - (t_c  + 1))}  \cdot R_j(\mathbf{S}_j(t)) }. 
  \label{eqn:foresighted_objective}
\end{equation}
Thus, the policy $\pi_j$ maps each state $\mathbf{S}_j(t) \in \mathbf{S}_j$ 
into an action, i.e., $\mathbf{A}_j(t) = \pi_j(\mathbf{S}_j(t))$, such that 
each action maximizes $R_j^{f}(\mathbf{S}_j(t_c))$. 

The policy can be deployed as a peer selection algorithm, which enables each peer 
to maximize its own long-term utility.
While the policy $\pi_j$ can be obtained using well-known methods such as value 
iteration and policy iteration~\cite{Bertsekas1976}, 
the environment 
dynamics keeps changing in practice,  
and thus, the policy needs to be updated frequently. This may 
require a high computational complexity.
Hence, it is important to reduce the complexity for finding the policy, 
such that the proposed algorithm can be efficiently deployed. 

\section{The Protocol Design }
\label{sec:design}
In this section, we describe the proposed protocol design that replaces the 
tit-for-tat and optimistic unchoke peer selection mechanisms, which are 
deployed in the regular BitTorrent systems, with the RL-based peer selection 
mechanism.

\begin{figure}
\centering
\includegraphics[width=9cm]{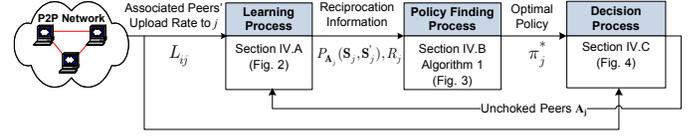}
\caption{Main processes in the proposed protocol design.} 
\label{fig:Processes_Flow}
\end{figure}

The protocol design is summarized in Fig.~\ref{fig:Processes_Flow}.
The protocol consists of three main processes running in parallel:
\begin{enumerate}
\item \emph{The learning process}, which provides updated information about 
statistical behaviors of the associated peers' resource reciprocation
$\mathbf{O}_j(t) ( \subseteq \mathbf{H}(t))$.
This process is necessary since the peers' reciprocation behaviors are not 
foretold. Therefore, peers are required to act in the environment in order to 
gain observation of the transition function and the rewards of the associated 
peers. 

\item \emph{The policy finding process}, which computes the policy using 
reinforcement learning. This process needs to be running 
in the entire downloading process as the changes of peers' 
reciprocation behaviors (identified by the learning process) 
can result in the policies obtained in the previous time slots being outdated.

\item \emph{The decision process}, which determines 
		the decisions on peer selection in each 
rechoke period based on the policy and the observed state. 
\end{enumerate}

More details about these processes are discussed next. 

\subsection{The Learning Process}
\label{subsec:TheLearningProcess}

It is difficult to estimate (or \emph{learn}) the other peers' states,   
rewards and state transition probabilities due to the unannounced information, 
network scalability constraints, time-varying network dynamics, etc. 
In our proposed protocol, a RL-based peer learns the other peers' states, 
rewards, state transition probability, etc., using the observations of its 
competing peers from the past. Thus, each peer needs to update the above 
information regularly through the learning process, while downloading content 
from its associated peers. 

\begin{figure}[!tbh]
		\centering
\includegraphics[width=8cm]{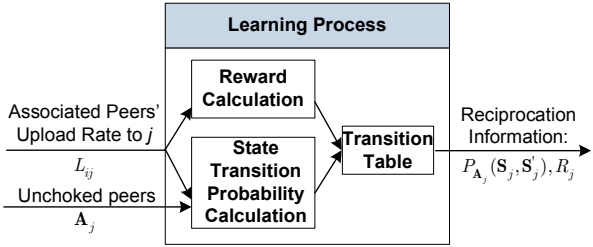}
\caption{The learning process.} 
\label{fig:LearningProcess}
\end{figure}

The learning process consists of two main methods that compute the estimated 
reward and state transition probability, which is depicted in 
Fig.~\ref{fig:LearningProcess}. 

\subsubsection{Computing Reward}

The reward of peer $j$ represents its total download rates from its associated 
peers estimated by peer $j$. In the rewards calculation method, the associated 
peers are classified into two types based on the available information about 
their resource reciprocation history.

For associated peers that have reciprocated their resources with peer $j$, 
referred to as \emph{peers with reciprocation history}, peer $j$ estimates 
their upload rates based on the weighted average of the past upload rate 
samples. This can reduce the fluctuation induced by the protocol and network 
dynamics in the sampled upload rates of the associated peers.
Specifically, peer $j$ estimates the upload rates $\hat L_{ij}$ of peer 
$i \in C_j$ based on recently observed resource reciprocation $L_{ij}$ as 
\begin{equation}
\hat L_{ij}(t+1) \leftarrow \alpha_j \cdot L_{ij}(t+1) + (1-\alpha_j) \hat L_{ij}(t)  
\label{eqn:EstimationUploadRate}
\end{equation}
where $\alpha_j$ denotes the weight for most recent resource reciprocation. 

For associated peers who have \emph{not} yet reciprocated their resources with 
peer $j$, which are referred to as \emph{peers without resource reciprocation 
history}, peer $j$ initializes the information about such peers by 
optimistically estimating that they reciprocate their resources with high 
probability and high upload rate. 
This enables peer $j$ to efficiently discover additional peers, and bootstrap 
newly joining peers, which is important for the efficiency of the system.
Whenever peer $j$ uploads to a peer without resource reciprocation history and 
the peer does not upload to $j$ in return, peer $j$ reduces the peer's presumed 
upload rate, as this provides $j$ with more confidence that the particular peer 
may not actively reciprocate its data. This also prevents the associated peers 
from taking advantage of a peer through optimistic initialization and possible 
free-riding. Note that white-washing~\cite{feldman04} is not possible in our 
design either, since peers are identified by their IP addresses.

\subsubsection{Finding State Transition Probability}
\label{subsubsec:finding_STP}
The state transition probabilities are updated every rechoke period, and thus, 
each peer can capture the time-varying resource reciprocation behaviors of its 
associated peers. Every rechoke period at $t+1$, peer $j$ stores 3-bit triplets 
for its associated peer $i$,
$\left(s_{ij}(t),a_{ji}(t), s_{ij}(t+1) \right)$.
Peer $j$ stores the triplets for its associated peers that are 
in its \emph{reduce peer set}, which will be discussed later in this 
section, or peers that uploaded to peer $j$ at time $t$ or $t+1$. 
In our design, we compute the state transition probability functions based on 
the empirical frequency, and assume that the state transition of each peer is 
independent. Thus, the state transition probability 
$P_{\mathbf{A}_j(t)} (\mathbf{S}_j(t), \mathbf{S}_j(t+1))$ from  
$\mathbf{S}_j(t) = (s_{1j}(t), \ldots, s_{Nj}(t))$  to 
$\mathbf{S}_j(t+1) = (s_{1j}(t+1), \ldots, s_{Nj}(t+1))$ given an action 
$\mathbf{A}_j (t) = (a_{j1} (t), \ldots, a_{jN}(t))$
can be expressed as 
\begin{equation}
		\notag
		P_{\mathbf{A}_j(t)} (\mathbf{S}_j(t), \mathbf{S}_j(t+1) ) = 
\prod_{i=1}^N \Pr(s_{ij}(t+1)| s_{ij}(t),a_{ji}(t)).
\end{equation}

\subsection{The Policy Finding Process}
\label{subsec:ThePolicyFindingProcess}

\begin{figure}
\centering
\includegraphics[width=8cm]{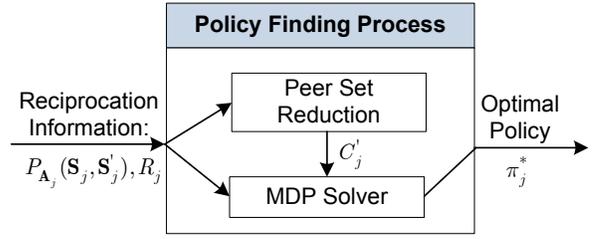}
\caption{A policy finding process.} 
\label{fig:PolicyFinderProcess}
\end{figure}

The policy finding process runs in parallel with the learning process, while
using the information obtained from the learning process. 
This process is depicted in Fig.~\ref{fig:PolicyFinderProcess}.
Finding the policy based on the reinforcement learning frequently may result in 
high computational complexity requirement, if the number of the associated 
peers becomes large. Hence, in order to practically implement the proposed 
algorithm, it is critical to reduce the number of peers that a peer considers 
for reciprocation (see Section \ref{sec:MDP}).
Therefore, this process needs to begin with reducing the set of associated 
peers, and then, finds the policy $\pi_j$ that maximizes the cumulative 
discounted expected reward (i.e., in Eq.~\ref{eqn:foresighted_objective}) in 
the reduced peer set. 

\subsubsection{Reducing Associated Peer Set}
As discussed in Section \ref{sec:MDP}, in order to find $\pi_j$ efficiently, 
it is important for peer $j$ to reduce the set of associated peers while 
selecting the peers who can reciprocate their resources with higher probability 
and with higher upload rate in the reduced peer set. Specifically, peer $j$ 
computes the expected rewards (or download rates) $\hat L_{ij}$ from each peer 
$i \in C_j$, defined as 
\begin{equation}
		\hat L_{ij}(t+1) = L_{ij}(t) \times \Pr(i \leadsto j),
  \label{eqn:ExpReward}
\end{equation}
where $\Pr(i \leadsto j)$ denotes the probability of resource reciprocation 
with peer $i$.
Based on the computed $\hat L_{ij}$, peer $j$ reduces its associated peer set 
by iteratively eliminating the peers with the smallest $\hat L_{ij}$ in its 
associated peer set. The algorithm for peer set reduction is presented in 
Algorithm~\ref{alg:PeerSetReduction}. 

\algsetup{indent=2em}
\begin{algorithm}[h!]
\caption{Peer-set Reduction Algorithm for Peer $j$}
\label {alg:PeerSetReduction}
\begin{algorithmic}[1]
\STATE {\bfseries INPUT :}
\\$\cdot$ $C_j$ - set of associated peers of peer $j$
\\$\cdot$ $T$ - targeted size of reduced peer set (constant)
\\$\cdot$ $L_{ij}$ - rewards (or download rates) from peer $i$
\\$\cdot$ $\Pr(s_{kj})$ - probability to be in $s_{kj}$
\\$\cdot$ $\Pr( i \leadsto j )$ - the resource reciprocation 
probability of peer $i$
\\$\cdot$ $c_1, c_2$ - constants such that $T \gg c_1 > c_2$ 
\STATE {\bfseries OUTPUT :} 
\\A reduced set of peers $C_j'\subseteq C_j$ where $|C_j'| = T$

\medskip

\FORALL{$i \in C_j$} \label{line:calcKstart}
\STATE $\hat L_{ij} = L_{ij}\times \Pr(i \leadsto j)$; \label{line:calcKend} 
\ENDFOR 
\STATE order $C_j$ in a non-decreasing order of the $\hat L_{ij}$; \label{line:orderI}
\STATE $C_j' \leftarrow  C_j$;\label{line:initI}
\WHILE{ $|C_j'| > T$ } \label{line:reduc1}
\STATE $G=\left\{ {C_j'}_1, \ldots, {C_j'}_{c_1} \right\}$;
\label{line:lowestK} 
\STATE calculate $\pi_{j,G}^{*}$ \footnotesize{\texttt{//policy for set
$G$}};\label{line:MDPsolve}
\FORALL{$k$ such that ${C_j'}_k \in G$}\label{line:calcP1} 
\STATE $\Pr (j \leadsto k) \leftarrow 0$ 
\footnotesize{\texttt{//estimate probability that $j$ unchokes $k$ based on
$\pi_{j,G}^{*}$}}
\FORALL{$s_{kj} \in \mathbf{S}_j$}
\IF{$\pi_{j,G}^{*}(s_{kj}) = 1$}
\STATE $\Pr(j \leadsto k) \leftarrow \Pr(j \leadsto k) + \Pr(s_{kj})$;
 \ENDIF
 \ENDFOR
  \ENDFOR \label{line:calcP2}
  \STATE order $G$ in a non-decreasing order of the $\Pr(j \leadsto k)$ values;
  \IF{$c_2 > |C_j'| - T$}
      \STATE $c_2 \leftarrow |C_j'|-T$;
  \ENDIF
  \STATE $C_j' \leftarrow C_j'- \left\{G_1, \ldots, G_{c_2} \right\}$; \label{line:removeC2}
 \ENDWHILE \label{line:reduc2}
 \RETURN $C_j'$
\end{algorithmic}
\end{algorithm}
The algorithm computes $\hat L_{ij}$ in \eqref{eqn:ExpReward} for $i \in C_j$ 
(lines \ref{line:calcKstart},\ref{line:calcKend}). Then, the associated peers 
are ordered based on the computed $\hat L_{ij}$ (line \ref{line:orderI}).
The peer set reduction is performed in the "while loop" 
(lines \ref{line:reduc1}-\ref{line:reduc2}) that reduces the peer set by $c_2$ 
peers in every iteration. In the loop, the algorithm selects $c_1$ peers with 
the smallest $\hat L_{ij}$ values denoted by $G$ (line \ref{line:lowestK}), 
from the reduced group of peers $C_j'$. It then obtains policy $\pi_{j,G}$ for 
the peers in $G$ (line \ref{line:MDPsolve}). Based on $\pi_{j,G}$, it 
calculates the probabilities for the peers to be unchoked 
(lines \ref{line:calcP1}-\ref{line:calcP2}). Given the calculated probability, 
it removes the $c_2$ peers with the lowest probability to be unchoked 
(line \ref{line:removeC2}). The algorithm runs until $|C_j'|=T$ 
(line \ref{line:reduc1}).

\subsubsection{Scaling}
Scaling of the rewards can be considered in the cases where the number of 
reciprocation samples is small in comparison to the difference between the 
lowest to the highest upload rates that are expressed in the P2P network.

\subsection{The Decision Process}
\label{subsec:TheDecisionProcess}

\begin{figure}
\centering
\includegraphics[width=8cm]{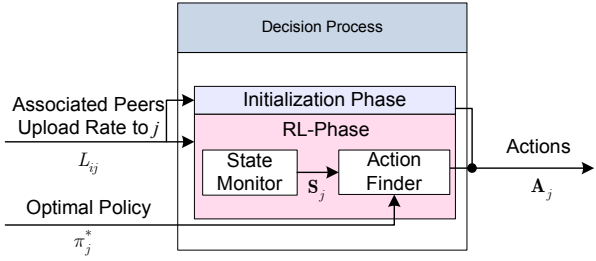}
\caption{A decision process.} 
\label{fig:DecisionProcess}
\end{figure}

The decision process includes two phases: the initialization phase and the RL
(Reinforcement Learning) 
phase, which is depicted in Fig.~\ref{fig:DecisionProcess}. 

\subsubsection{Initialization Phase} 
Since no information about associated peers is available for a newly joining 
peer $j$, peer $j$ begins with adopting the regular BitTorrent mechanisms 
(i.e., the tit-for-tat mechanism and the optimistic unchoke mechanism) in the 
initialization phase. This enables the peer to collect information such as the 
rewards and state transition probabilities with respect to its associated peers. 
During this phase, $j$ discovers new peers, i.e., downloads from peers for the 
first time. 
Once $j$'s peer discovery is slowed down (see Section~\ref{sec:implementation} 
for more details), it replaces the regular BitTorrent RL-based peer selection 
mechanisms, and operates in the RL phase. 

\subsubsection{RL Phase}
In this phase, peer $j$ determines the decisions on peer selection based on the 
policy obtained from the policy finding process in every 
rechoke period. 
Peer $j$ determines its current state $\mathbf{S}_j$ and the corresponding 
action $\mathbf{A}_j$ based on the policy $\pi_j$, i.e., 
$\mathbf{A}_j = \pi_j (\mathbf{S}_j)$. Note that $\mathbf{A}_j$ is a set of decisions
on peer selection of peer $j$, i.e. either to choke or to unchoke.

\section{Implementation}
\label{sec:implementation}
In this section, we discuss our proposed protocol prototype and study how to 
determine several design parameters. 

Our RL-based client is implemented on top of the \emph{Enhanced CTorrent} 
client, version 3.2~\cite{ctorrent}. 
We enhance the original client such that our client can operate in
\textit{RL-enhanced mode}, where it reciprocates its resources using the 
proposed RL-based mechanism, or in 
\textit{regular mode}, where it reciprocates its resources based on the regular 
BitTorrent peer selection mechanism.
We add the functionality for the RL-enhanced mode to support the proposed 
protocol requests. More specifically, in the RL-enhanced mode we implemented 
the three different processes that are discussed in Section~\ref{sec:design}.

\subsection{The Learning Process}
The learning process consists of two methods, the reward calculation method 
and the state transition probability calculation method. In 
Section~\ref{subsubsec:finding_STP} we discussed how to estimate the state 
transition probability, and in this section we will describe the reward 
calculation method.

The reward calculation method can be applied differently depending on the 
associated peer types: peers with or without reciprocation history. 

\subsubsection{Peers with Reciprocation History}
\begin{figure}[!tb]
\centering 
\includegraphics[width=8cm]{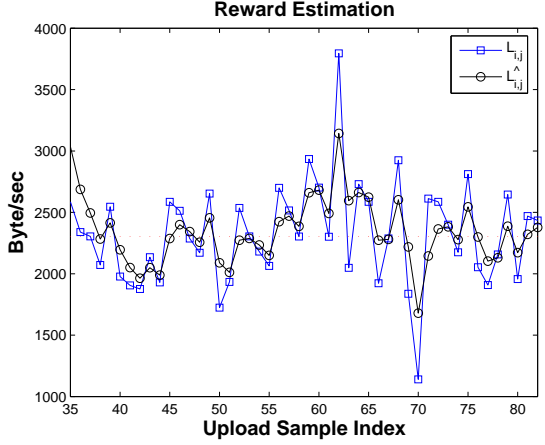}
\caption{Upload samples and reward estimations} 
\label{fig:uploadEst}
\end{figure}

While calculating the reward of a peer with resource reciprocation history, 
the samples of $L_{ij}$ will obviously fluctuate over the rechoke time period 
due to the experienced P2P network dynamics. 
Because of this fluctuation, $L_{ij}$ samples may be atypical. Thus, 
a typical upload rate of a peer with reciprocation history can be estimated 
based on a weighted average of the samples as in 
\eqref{eqn:EstimationUploadRate}. 
This is the estimated reward of peer $j$ obtained from peer $i$. 
As recent resource reciprocations are considered more important than the past 
reciprocations, we set $\alpha_j \geq 0.5$.  
Based on several trials for $\alpha_j$ such that  
$0.5+\epsilon \leq \alpha_j \leq 1-\epsilon$ for small $\epsilon \geq 0$ on 
various sets of our experiments (see more details in 
Section~\ref{sec:evaluation}), we can verify that a smaller $\alpha_j$ achieves 
less fluctuation of the reward. Thus, we set $\alpha_j$ as $0.5$.
Fig.~\ref{fig:uploadEst} shows an example for sampled upload rates $L_{ij}$ of 
a peer $i$ in our network and the correspondingly estimated upload 
rates $\hat L_{ij}$ measured by another associated peer in the network. 
We can clearly observe less variations of the $L_{ij}$ in the computation 
of the $\hat L_{ij}$. 

\subsubsection{Peers without Reciprocation History}
\begin{figure}[!tb]
\centering
\includegraphics[width=8cm]{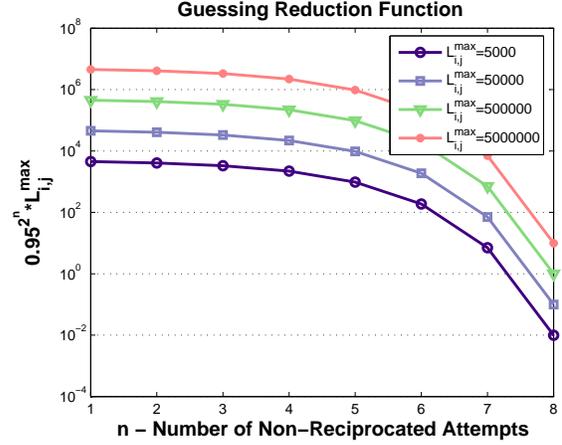}
\caption{Guessing reduction function} 
\label{fig:conjectureFunc}
\end{figure}

If there is no resource reciprocation history for peer $i$, a leecher $j$ 
optimistically initializes the information about the rewards and the 
reciprocation probabilities of its associated peers. 
Specifically, the initial estimated upload rate is set to be the highest 
upload rate $L_{ij}^{max}$ that is pre-determined in the P2P network, i.e., 
$L_{ij} \leftarrow L_{ij}^{max}$, and the probability of reciprocation 
with $j$ is initiated to $1$, i.e., $\Pr(i \leadsto j) \leftarrow 1$.
This optimistic initialization enables newly joining leechers to download almost
immediately. Peer $j$ needs to continue updating the initially assumed reward 
in every non-reciprocated event (i.e., peer $j$ uploads resources to peer $i$ 
while peer $i$ does not upload resources to peer $j$). When peer $j$ estimates 
the reward for peer $i$, peer $j$ can assume that 
\begin{itemize}
		\item [(i)] $\hat L_{ij}$ satisfies  
\begin{equation}
		\frac{\hat L_{ij} (n-1)}{\hat L_{ij} (n)}  < \frac{\hat L_{ij}
		(n)}{\hat L_{ij} (n+1)}
\end{equation}
where $n$ denotes the number of non-reciprocated events,
\item  [(ii)] $\hat L_{ij} (n)$ decreases exponentially such that it 
approaches 0 after several attempts. 
\end{itemize}
The assumption (i) means that the ratio of the estimated rate of two 
consecutive events is an increasing function of $n$. This also implies that the 
increasing uncertainty about peer $i$'s reciprocation behavior. Moreover, the 
assumption (ii) is required to prevent the non-reciprocated behavior including 
free-riding. Thus, a function satisfying (i) and (ii) can have a form, such as 
\begin{equation}
		f(n) = \beta^{g(n)}\times L_{ij}^{max}
\end{equation}
where $\beta (<1)$ is a constant and $g(n)>1, \forall n\geq 1$ is a function 
that grows faster than a linear function. In our implementation, we use 
function $f(n) = 0.95^{2^n}\times L_{ij}^{max}$ because the function satisfies 
properties (i) and (ii), as shown in Fig.~\ref{fig:conjectureFunc}. 

\subsection{The Policy Finding Process} 
As shown in Section~\ref{sec:design}, in every iteration of the policy finding 
process, the associated peer set is first reduced. Based on our experiments, we 
observe that when the reduced size of peer set is more than 7 peers, finding 
the RL-based policy slows down the RL-enhanced client performance.
Thus, in our implementation, we set the size of the reduced peer set as seven, 
i.e., $T=7$ in Algorithm~\ref{alg:PeerSetReduction}. 

The computed policy holds for up to additional three rechoke periods, 
which is determined considering the tradeoff between the time for enough 
reciprocation and the time for capturing the network dynamics.  

\subsection{The Decision Process}
The initialization phase and the RL phase in the decision process
are implemented as follows. 

\subsubsection{Initialization Phase}  
In the initialization phase, peer $j$ makes its decisions on peer selection 
based on the regular BitTorrent mechanisms, as it does not have enough 
information to calculate the policy. 

\begin{figure}[!tb]
		\centering
		\includegraphics[width=8cm]{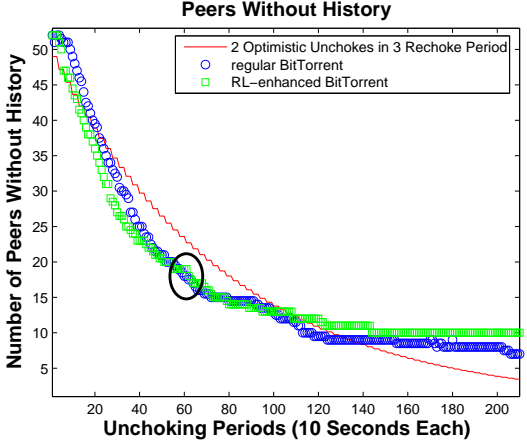}
\caption{Discovering rate of peers without history} 
\label{fig:DiscoveringRatio}
\end{figure}

In order to determine the duration of the initialization phase we study 
extensive experiment results, which include both flash crowd scenarios as well 
as steady state scenarios. In these experiments, the number of peers that have 
not uploaded to peer $j$ from the beginning of the downloading process is 
counted every rechoke period. Fig.~\ref{fig:DiscoveringRatio} shows the median 
of the counted numbers of peers collected from all the leechers in the network 
over time (rechoke periods) for several experiments of flash-crowd scenarios.

Fig.~\ref{fig:DiscoveringRatio} shows that the peer counted value is 
exponentially decreasing and stabilized quickly. Then, peer $j$ can switch from 
the initialization phase to the RL phase. 
In our implementation, a peer $j$ counts the number of peers without 
reciprocation history within every rechoke period. Once the count reduces by 
one in duration of three rechoke periods and for two consecutive durations 
(i.e., six rechoke periods), peer $j$ switches to the continuous phase and 
begins to adopt the RL-based strategy.
Based on our experiments, peers switch from the initialization phase to the 
continuous phase approximately 60 rechoke periods later in the flash-crowd 
scenarios and approximately 36 rechoke periods later in the steady-state 
scenarios. However, different network settings might lead to different 
durations of the initialization phase.

\subsubsection{RL Phase} 
In the RL phase, the peer selection decisions are made based on the RL policy 
every ten seconds (as in regular BitTorrent). The selected peers will be 
unchoked for a rechoke period of ten seconds. The minimum number of unchoked 
peers is four. 
The number of unchoked peers can increase if
\begin{itemize}
\item [(1)] The peer that makes the peer selection decision does not saturate
its upload capacity, or,
\item [(2)] The upload bandwidth of the peer that makes the peer selection
decision is higher in comparison to most of the peers it interacts with. 
\end{itemize}

We compare the performance of the proposed protocol with that of the regular 
BitTorrent implemented in the Enhanced CTorrent client. The minimum number of 
unchocked slots in the regular BitTorrent implementation is also set as four. 
The number of slots can increase if a peer's upload capacity is not saturated. 
In this implementation, one unchoke slot is always reserved for optimistic 
unchokes that are rotated every three rechoke periods.

\section{Experimental Evaluation}
\label{sec:evaluation}
We perform extensive experiments on a controlled testbed, in order to evaluate 
the properties of the proposed protocol. 

\subsection{Methodology}
\label{sec:methodology}
All of our experiments are performed on the Planetlab experimental 
platform~\cite{planetlab04}, which utilizes the nodes (machines) located across 
the globe.
We execute all the experiments consecutively in time on the same set of 
nodes. Unless otherwise specified, the default implementations
of leecher and seed in regular BitTorrent systems are deployed. 

The upload capacities of the nodes are artificially set according to the 
bandwidth distribution of typical BitTorrent leechers~\cite{piatek07}.
The distribution was estimated based on empirical measurements of BitTorrent 
swarms including more than 300,000 unique BitTorrent IPs.
Since several nodes are incapable to match the target upload capacities 
determined by the bandwidth distribution, 
we scale the upload capacity and other relevant experimental parameters such 
as file size by 1/20th. 
However, we have not set limitations on download bandwidth.

All peers begin the download process simultaneously, which emulates a flash 
crowd scenario. 
The initial seeds have stayed connected through out the entire experiment.
To provide synthetic churn with constant capacity, leechers disconnect 
immediately after completion of downloading the entire video file, and
reconnect as new comers immediately while requesting the entire video file 
again. This enables our experiments to have the same upload bandwidth 
distribution for the duration of the experiment.

Unless otherwise specified, our experiments host 104 Planetlab nodes, 100 
leechers and 4 seeds with a combined capacity of 128 KB/s, serving a 99 MB video 
file. 

\subsection{Experiment Results: Single RL Leecher in a Network}
\begin{figure}[t]
\centering
\includegraphics[width=0.8\columnwidth]{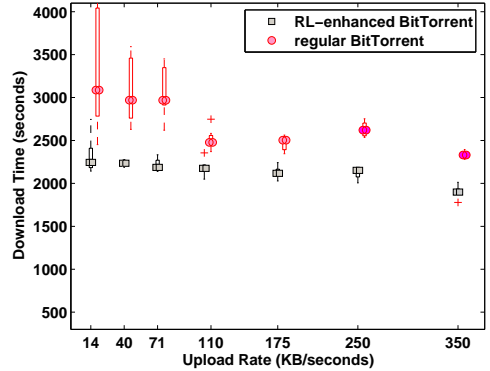}
\caption{Leecher download time} 
\label{fig:single}
\end{figure}

We start with the experiment where only a single leecher adopts the 
\emph{RL-enhanced} protocol, while the rest of the leechers in the network run with 
the regular BitTorrent, and there are no \FR s in the network (note that this 
is a common scenario that was tested by other proposed protocols such 
as~\cite{levin08, piatek07}). 
Fig.~\ref{fig:single} compares the download time of a single 
leecher, while adopting the \emph{RL-enhanced} protocol and the regular BitTorrent 
protocol as a function of the leecher's upload capacity over 7 trials.

In Fig.~\ref{fig:single}, as in~\cite{mcgill78} separate boxplots are depicted 
for the different scenarios. The top and the bottom of the boxes represent the 
75th and the 25th percentile sample of download time, respectively, over all 7 
runs of the experiments. The markers inside the boxes represent the median, 
while the vertical lines extending above and below the boxes represent the 
maximum and minimum of samples of download time within the ranges of 1.5 time 
the box height from the box boarder. Outliers are marked individually with 
``$+$" mark.

The results in Fig.~\ref{fig:single} provide several insights into the 
operation of our RL-based proposed protocol.
High and Low capacity leechers benefit from the \emph{RL-enhanced} with 12\%-27\%
improvement of their download time performance as indicated by the median.
This improvement provides leechers with an incentive to adopt the proposed 
protocol. Moreover, the RL-based strategy does not simply improve 
performance; it also provides more consistent performance across multiple 
trials. By selecting to unchoke peers based on historical behavior information, 
our proposed protocol avoids the randomization present in the regular BitTorrent 
tit-for-tat and optimistic unchoke implementations, which cause to unstable 
peer selections and results in slow convergence.\\
%
%
\\{\bfseries Peer Selection Mechanism Stability}
\begin{figure}[t]
\centering
\includegraphics[width=0.8\columnwidth]{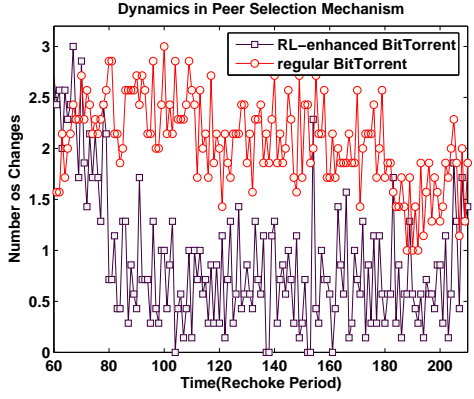}
\caption{Peer selection mechanism dynamics} 
\label{fig:dynamics}
\end{figure}
We further study the peer selection mechanism stability.
The stability of the peer selection mechanism affects directly the performance 
of the system since once a peer starts to upload to another peer it takes time 
till the peer reaches its full capacity. In the BitTorrent 
protocol~\cite{cohen03} the author suggests allowing 30 seconds for a peer to 
reach its full capacity. 
Thus, a system that has a high fluctuation in peer selection will have many 
occurrences of peers that do not reach their full capacity.

We compare the peer selection fluctuations of the two protocols.
A stable peer selection mechanism should minimize the peer selection 
fluctuations. We measured peer selection fluctuations by comparing the peer 
selection decisions during two consecutive rechoke periods and measuring the 
difference between the two decisions, e.g., replacing an unchoked peer by a 
different peer counts as one change.
Fig.~\ref{fig:dynamics} indicates the average number of peer selection 
changes as a function of time (rechoke period units) for a single peer. 
It shows that the average number of peer selection changes is lower in the 
\emph{RL-enhanced} network for the majority of the time, with an average of 2.1 
changes in the regular \BT~network as compared to 0.9 average changes in the 
\emph{RL-enhanced} network. Thus, the \emph{RL-enhanced} peer selection 
mechanism is more stable than the peer selection mechanism in the regular \BT, reducing the peer selection fluctuations by an average of 57\%.

Note that the optimistic unchoke mechanism contributes about 1 change 
every 3 rechoke periods, thus contributing an average change of about 
$\frac{1}{3}$ per time unit in the regular \BT~network. Therefore, the decrease 
in optimistic unchokes is not the main reason for the stability improvement of 
the peer selection mechanism. 
Instead, replacing the tit-for-tat mechanism, which relies on short-term 
history of associated peers with the RL-based mechanism that 
relies on a long history and performs foresighted unchoking decisions is the 
main contributor for this stability.

\subsection{Experiment Results: Performance of Leechers in Network without
Free-Riders}

\begin{figure}[t]
\centering
\includegraphics[width=0.8\columnwidth]{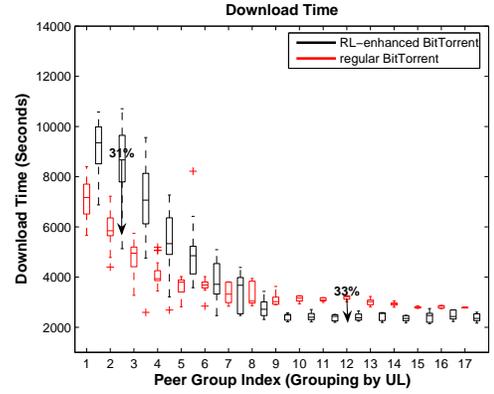}
\caption{Download completion time for leechers.} 
\label{fig:DLtime}
\end{figure}

We compare a system consisting of all leechers adopting the regular BitTorrent 
protocol, to a system consisting of all leechers running in \emph{RL-enhanced} mode, adopting the RL-based strategy.  
In this section, we assume that there are no free-riders in the network. 
Note that this experiment hosted only 50 leechers. 
Fig.~\ref{fig:DLtime} shows the download completion time of leechers.
For each group of leechers having the same upload capacity, separate boxplots 
are depicted 
for the different scenarios.

The results show the clear performance difference among high-capacity leechers, 
which are the fastest 20\% leechers, and low-capacity leechers, which are the 
slowest 80\% leechers. 
High-capacity leechers can significantly improve their download
completion time -- leechers having the upload capacity of at least 18kB/sec 
improve their download completion time by up to 33\% in median. 
Unlike in the regular BitTorrent system, where leechers determine their peer 
selection decisions based on the myopic tit-for-tat that uses only the last 
reciprocation history, the \emph{RL-enhanced} leechers determine their peer 
selection decisions based on the long term history. This enables the leechers 
to estimate the behaviors of their associated peers more accurately. Moreover, 
since part of the peer selection decisions is randomly determined in the 
regular BitTorrent, there is a high probability that high capacity leechers 
need to reciprocate with the low-capacity leechers~\cite{piatek07}. 
However, the randomly determined peer selection decisions are significantly 
reduced in the proposed approach, as the random decisions are taken only in the 
initialization phase or in order to collect the reciprocation history of newly 
joined peers. 
As a result, the high capacity leechers increase theie probability to 
reciprocate resources with other high capacity leechers.

\begin{figure}[t]
\centering
\includegraphics[width=0.8\columnwidth]{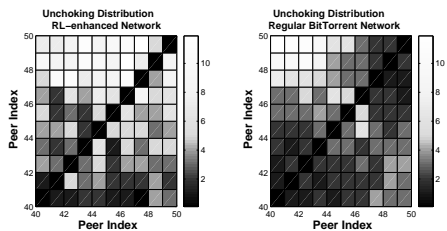}
\caption{Unchokes among the $20\%$ fastest peers.} 
\label{fig:Unchoke_dist}
\end{figure}

This is confirmed in the results of Fig.~\ref{fig:Unchoke_dist}, which shows 
the unchoking percentage among the $20\%$ high capacity leechers, 
comparing the two different systems.
It is clearly observed that the collaboration among high capacity leechers 
improves when leechers adopt the RL-based strategy. 
Thus, we can conclude that the RL-based strategy improves the 
incentive mechanisms in BitTorrent networks: as a leecher contributes more to 
the network, it achieves higher download rate. 
 
\begin{figure}[t]
\centering
\includegraphics[width=0.8\columnwidth]{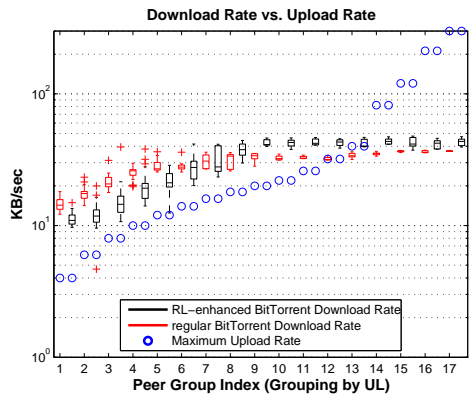}
\caption{Download rates versus upload rates.} 
\label{fig:dl_vs_ul}
\end{figure}

Recent studies~\cite{piatek07,Buddies,bharambe06,guo05} show that 
the regular BitTorrent protocol suffers from unfairness particularly for high 
capacity leechers.
In Fig.~\ref{fig:dl_vs_ul}, we compare the upload rates and the average 
download rates of the leechers. The ratio of these values can indicate the 
degree of fairness in the system.
The results in Fig.~\ref{fig:dl_vs_ul} show that fairness is improved in 
the \emph{RL-enhanced} network, since high-capacity leechers increased their 
download rate getting closer to their upload rate, in spite of the restriction 
of limited seeds' upload rate. On the other hand, in the \emph{RL-enhanced} network, 
the download rates of low-capacity leechers decrease, getting close to their 
upload rates by at most $36\%$, compared to the regular BitTorrent system.
However, all the peers that are slowed down by the RL-based 
strategy still  download faster than their upload rate. \\

\subsection{Experiment Results: Performance of Leechers in Network with
Free-Riders}

\begin{figure}[t]
\centering
\includegraphics[width=0.8\columnwidth]{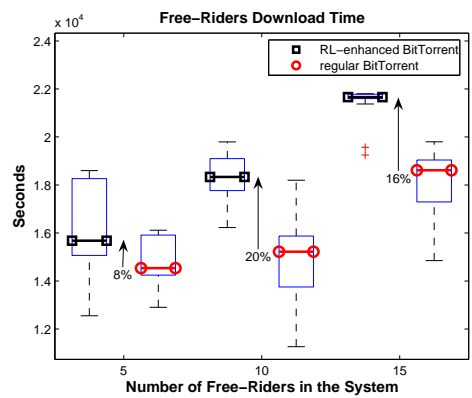}
\caption{Download completion time for free-riders.} 
\label{fig:FR_DLtime}
\end{figure}

In this section, we investigate how effectively the proposed protocol can 
prevent selfish behaviors such as free-ridings. Note that the RL-based strategy shows a similar performance for the leechers that upload 
their content in a network that includes free-riders (i.e., shows the 
improved fairness, etc.). Hence, in this section, our focus is on studying how 
the free-riders are punished due to their selfish behaviors. 
Fig.~\ref{fig:FR_DLtime} shows the time that the free-riders complete 
downloading 99MB video file in a network consisting of 50 contributing 
leechers, and increasing number of free-riders (i.e., 5, 10, and 15 
free-riders). It compares the results of the \emph{RL-enhanced} network to the 
regular BitTorrent network. Fig.~\ref{fig:FR_DLtime} confirms that in the 
\emph{RL-enhanced} network the leechers are able to effectively penalize the 
free-riders, as it takes longer time for the free-riders to complete their 
downloads (requires 8\%-20\% more time as measured by the median, in 
comparison to the regular BitTorrent protocol).

\begin{figure}[t]
\centering
\includegraphics[width=0.8\columnwidth]{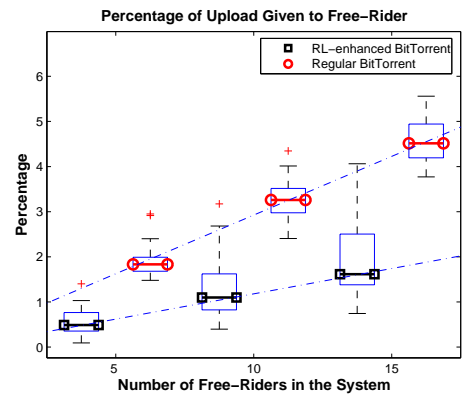}
\caption{Percentage of free-riders' downloads from contributing leechers.} 
\label{fig:upload_to_fr}
\end{figure}

The \emph{RL-enhanced} leechers can efficiently capture the selfish behaviors of 
the free-riders. Thus, they unchoke the free-riders with a significantly lower 
probability. Hence, the free-riders can download their content mainly from 
seeds and not from the leechers. The results shown in 
Fig.~\ref{fig:upload_to_fr} also confirm that the leechers in the regular 
BitTorrent network upload approximately 2.8-3.7 times more data to the 
free-riders compared to the \emph{RL-enhanced} network.
This also shows that the \emph{RL-enhanced} networks are more robust to the selfish 
behaviors of peers than the networks operating with the regular BitTorrent 
protocol. For example, in the network with 15 free-riders, the leechers in the 
regular BitTorrent systems upload 4.5\% of their total upload capacity to 
free-riders, while they only upload 1.6\% of their total upload capacity in the 
\emph{RL-enhanced} network. Thus reducing by 64\% their upload capacity to \FR s.

Therefore, our experiment results confirm that the RL-based strategy provides incentives for adoption because it improves the peer's download rate, improves 
the stability of the peer selection mechanism, improves collaboration among 
high capacity peers, improves fairness in the system, and discourages 
non-cooperative behaviors such as free-riding.

\section{Related Work}
\label{sec:related}

A fairly large number of P2P architectures that support distribution of 
multimedia over the Internet has been proposed in the last years within the 
scientific community~\cite{chu01,Banerjee02, zigzag, Castro03,PALS}. 
More specifically, BitTorrent, the protocol that dominates the traffic on the 
Internet~\cite{ipoque}, has been highly influential in the design and 
development of many other modern commercial P2P streaming systems such 
as~\cite{PPLive, UUSEE, coolstreaming}. 

Extensive research has focused on modeling and analyzing the performance of 
the BitTorrent systems, since the main mechanisms and the design rationale of 
the BitTorrent protocol were first described~\cite{cohen03}.

Qiu and Srikant~\cite{qiu04} studied a fluid analytical model of BitTorrent 
systems. They analytically studied the choking mechanism and how it affects the 
peer performance. They showed that the optimistic unchoke mechanism may allow 
free-riding. They also claimed that the system with tit-for-tat strategy 
eventually converges with a Nash equilibrium where fairness is achieved and all 
peers download at their upload capacities. 
However, as shown in our results, which are in consistent with other existing 
works such as~\cite{piatek07,Buddies,guo05,levin08} the choking mechanism in 
BitTorrent may fail to attain fairness for realistic swarms. 
Fan \textit{et al.}~\cite{fan06} characterized the design space of 
BitTorrent-like protocols capturing the fundamental tradeoff between 
performance and fairness. We also study such tradeoffs and show that the 
RL-based strategy improves the fairness in the system for the 
cost of reduced download rates of low-capacity leechers. This encourages 
leechers to contribute more resources (i.e., maximize their upload rate). 
Levin~\textit{et al.}~\cite{levin08} propose an auction base model to model the 
peer selection mechanism, claiming that BitTorrent uses auction to decide which 
peers to unchoke and not the tit-for-tat as widely believed.

Other researchers have studied the feasibility of free-riding behavior; 
Shneidman \textit{et al.}\cite{shneidman03} showed that it is possible to 
free-ride in BitTorrent systems. They identified forms of strategic 
manipulation that are based on Sybil attacks and uploading garbage data.
Liogkas \textit{et al.}~\cite{liogkas06} implemented three exploits that allow 
free-riders to obtain higher download rates under specific circumstances. 
Locher \textit{et al.}~\cite{locher06} with BitThief extended this work by 
showing that free-riders can achieve higher download rate, even in the absence 
of seeds. 
Similarly, Sirivianos \textit{et al.}~\cite{sirivianos07} showed that 
a free-rider, which can maintain a larger-than-normal view of the system, has 
a much higher probability to receive data from seeds and via optimistic 
unchoke. Our protocol replaces the optimistic unchokes, the most important 
vulnerability identified in these studies, with the RL-based 
policy based unchokes.

Fairness in BitTorrent systems was studied as well. 
Geo \textit{et al.}~\cite{guo05} showed the lack of fairness in BitTorrent 
systems. Piatek \textit{et al.}~\cite{piatek07}, observed the presence of 
significant altruism in BitTorrent, where peers make contributions that do not 
directly improve their performance. Izhak-Ratzin in~\cite{Buddies} identified 
the potential lack of fairness and proposed the Buddy protocol that matches 
peers with similar bandwidth. Legout \textit{et al.}~\cite{legout07} studied 
clustering of peers having similar upload bandwidth. They observed that when 
the seed is under provisioned; all peers tend to complete their downloads 
approximately at the same time, regardless of their upload rates. Moreover, 
high-capacity peers assist the seed to disseminate data to low-capacity peers. 
This can happen because the tit-for-tat strategy is based on short-term history.
 A peer can benefit from the tit-for-tat strategy only if it can continuously 
upload pieces and as long as it receives pieces of interest in return. Piatek 
\textit{et al.}~\cite{piatek08} showed that this is not always possible, as 
peers can have no piece to offer. Our work also considers the unfairness in 
BitTorrent systems, and shows that the proposed approach can improve the 
fairness by using a long-term history based strategy.

In order to reduce free-riding and encourage collaboration, various reputation 
systems have been proposed. Payment systems (e.g.,\cite{wil02, karma}), which 
enable peers to earn credits according to their uploads to other peers have 
been proposed. However, in practice these systems require a centralized entity 
to prevent cheating, and thus, have arguably scalability limitations.
To overcome such weaknesses in payment systems, various designs of reputation 
systems have been proposed 
(e.g.,\cite{buchegger04, xiong04, yang05, lian06, piatek08}). 
In these systems, peers can make choking decisions bases on private history as 
well as globally shared history. However, these reputation systems require 
significant communication overheads to maintain the global history. Moreover, 
there is no guarantee that each peer expresses the same behavior to different 
peers with different attributes.
%

Other researchers have also acknowledged the importance of contribution 
incentives in P2P systems and have proposed different alternatives.
Anagnostakis \textit{et al.}~\cite{Anagnostakis04} suggested to extend the 
BitTorrent incentives to \textit{n-way} exchanges among rings of peers, 
providing incentive to cooperate. 
Piatek \textit{et al.}~\cite{piatek07} proposed the BitTyrant client, 
who applies a new peer selection mechanism that reallocates upload 
bandwidth to maximize peers' download rates. 
However, whereas the appearance of a single BitTyrant client in a BitTorrent 
system reveals improving performance; in the case of a widespread adoption the 
system performs a severe loss of efficiency~\cite{Carra08}.
Levin~\textit{et al.}~\cite{levin08} proposed the propshare client that 
rewards other peers with proportional shares of bandwidth. They show that the 
propshare client improves performance in a swarm consisting predominately of 
BitTorrent peers. However, when the majority of peers run with propshare 
clients there is no clear difference in performance in comparison to the 
regular \BT~protocol. 

In addition, all these systems relay on short-term history aim to maximize the 
immediate utility but not the long-term utility, which can show only suboptimal 
performance. To the best of our knowledge, we are the first to propose the 
RL-based strategy that can replace the existing mechanisms 
deployed in BitTorrent protocol, while maximizing long-term utility of 
participating leechers.   

\section{Conclusion}
\label{sec:conclusion}

In this paper, we propose a BitTorrent-like protocol that replaces the 
tit-for-tat and the optimistic unchoke peer selection mechanisms in the regular 
BitTorrent protocol with a novel RL-based mechanism.

In our proposed protocol the evolution of the peers' interactions across the 
various rechoke periods are modeled as repeated interactions in a game. During 
the repeated multi-peer interactions, the peers can observe partial historical 
information of associated peers' reciprocation behaviors. Through this the peers 
can estimate the impact on their future rewards and then adopt their best peer 
selection action. The estimation of the impact on the expected future reward is 
performed using reinforcement-learning, as it allows the peers to improve their 
peer selection mechanism using only knowledge of their own past interactions, 
without knowing the complete reciprocation behavior of the peers in the 
network. 

Our experiment results show that our proposed protocol improves the stability 
of the peer selection mechanism, improves collaboration among high capacity 
peers, improves fairness in the system, enhances the robustness of the network 
by effectively discouraging non-cooperative behaviors such as free-riding, and 
importantly improves the downloading rates of the peers deploying the protocol.

\bibliographystyle{IEEEtran}
\bibliography{IEEEabrv,RL_P2P}

\end{document}